\documentclass[english,showkeys]{revtex4}
\usepackage[T1]{fontenc}
\usepackage[latin1]{inputenc}
\usepackage{graphicx}
\usepackage{amssymb}

\makeatletter

\usepackage{amsfonts}

\usepackage{babel}
\makeatother
\begin{document}

\title{Robustness of planar random graphs to targeted attacks}

\author{J.-P. Kownacki}

\email{kownacki@ptm.u-cergy.fr}

\affiliation{Laboratoire de Physique Théorique et Modélisation, CNRS-Université
de Cergy-Pontoise - UMR 8089 , 2 avenue Adolphe Chauvin, 95302 Cergy-Pontoise
Cedex, France}

\begin{abstract}
In this paper, robustness of planar trivalent random graphs to targeted
attacks of highest connected nodes is investigated using numerical
simulations. It is shown that these graphs are relatively robust.
The nonrandom node removal process of targeted attacks is also investigated
as a special case of non-uniform site percolation. Critical exponents
are calculated by measuring various properties of the distribution
of percolation clusters. They are found to be roughly compatible with
critical exponents of uniform percolation on these graphs. 
\end{abstract}



\maketitle

\section{Introduction}

Robustness or fragility of a graph characterize its behavior when
systematic or random deletion of a fraction of nodes is performed.
These questions are of practical importance for real complex networks
(World Wide Web, social networks, cells,etc) and, recently, there
has been much interest in investigating targeted and random attacks
for two families of random complex networks \cite{key-1,key-2,key-3},
namely exponential and scale-free networks \cite{key-3BIS} . Resilience to random
deletion of nodes can, for example, be seen as tolerance against error,
\textit{i.e.} random failure of a fraction of nodes. Targeted deletion
of high degree nodes can simulate attacks of hackers on computer networks.
However, besides these practical issues, investigating random and
targeted attacks on abstract graphs is of great theoretical interest
in its own right. It has to be noted that the random attack problem is equivalent
to uniform site percolation \cite{key-3TER}.

In this paper, we study a family of purely abstract random graphs
extensively used in the past decades as discrete models for euclidean
quantum gravity \cite{key-4}, namely random $\Phi^{3}$ planar graphs.
These planar graphs are made of trivalent vertices and look locally
like the regular honeycomb lattice. But the faces are not necessarily
hexagonal, so that long distance properties are very different from
those of honeycomb lattices. Note that site percolation on these graphs
has already been investigated \cite{key-5,key-6}. By means of Monte-Carlo
simulations, we study resilience of these graphs against targeted
attacks defined as a systematic face removal process, {\it i.e.} removal
of all links and vertices that belong to the edges of the targeted face. More precisely, 
we investigate connected clusters distributions
when all faces whose number of edges is larger than a cut-off $k_{max}$
are removed. 

A key feature in the behavior of graphs against targeted attacks is
the degree distribution. However, vertex coordination numbers on
complex networks are uncorrelated random variables so that
the degree distribution is sufficient to characterize these random graphs.
On the contrary, planarity constraints induce long range correlations for 
random $\Phi^{3}$ planar graphs so that the degree distribution is not enough
to characterize these graphs \cite{key-6BIS}. This characteristic is also shared by 
Vorono\"{\i}/Delaunay graphs. However, correlations in the latter case 
decrease much faster than for random $\Phi^{3}$ planar graphs. This fundamental
difference between complex networks and random $\Phi^{3}$ planar graphs induces
radically different behaviors under uniform percolations problems. The main purpose of this
work is then to determine whether it also induces different behaviors against non-uniform
percolation like targeted attacks, and, if so, to quantify these differences. 

It has to be noted that, even if random $\Phi^{3}$ planar graphs have been used 
in Euclidean quantum gravity, the problem studied here is not relevant to
these issues. Indeed, there is no back-reaction of any degree of freedom
on the connectivity of the graphs. In other words, the disorder due to random degree 
distribution is quenched and we study the average behavior against targeted
attacks of a graph picked at random in the ensemble of  random $\Phi^{3}$
planar graphs. However, this work could be the starting point of the study of 
a modified model of quantum gravity characterized by
matter fields (Ising spin, for instance) coupled to the ensemble of graphs
obtained from the random $\Phi^{3}$ planar graphs after a systematic
node delation procedure.

Besides the size of the largest connected component $S$, we study
several properties of connected cluster of graphs for various values
of $k_{max}$, in particular, the density of the second largest cluster
$s_{2}$ and the average size of finite clusters $S$. Considering
$k_{max}$ as a continuous parameter, it is possible to extract a
critical parameter $k_{c}$ similar to a percolation threshold, and
to compute critical exponents $\beta$, $\gamma$ and $\nu$ associated
to the second largest cluster, average cluster size and correlation
length. These exponents are compared with those of uniform percolation
problem on random $\Phi^{3}$ planar graphs.

This paper is organized as follows: in section \ref{sec:planar_graphs},
we introduce the model of planar $\Phi^{3}$ random graphs; in section
\ref{sec:Targeted}, targeted attacks are defined and results of numerical
simulations are reported; the problem is described as non-uniform
percolation and critical exponents are extracted from numerical results
in section \ref{sec:Non-uniform}; section \ref{sec:Conclusions}
contains conclusions and perspectives.

\section{Random $\Phi^{3}$ planar graphs\label{sec:planar_graphs}}

\begin{figure}
\includegraphics[clip,width=0.4\textwidth,keepaspectratio]{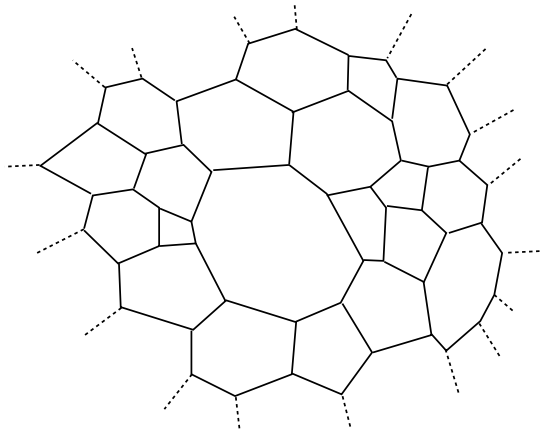}

\caption{A $\Phi^3$ planar graph }

\label{phi3} 
\end{figure}

We consider the class of trivalent planar graphs, \textit{i.e} all
graphs without boundaries that can be drawn on a sphere and where
each vertex is connected to exactly three neighbors. Moreover, two
distinct vertices are connected by at most one link and no vertex
is connected to itself. An example of $\Phi^3$ graph is shown in figure
\ref{phi3}.

 For a given number of vertices $N$, all graphs
can be obtained by gluing together $N$ trivalent vertices in all
possible ways satisfying the strong constraint of planarity. The Euler
characteristic $\chi=N-N_{l}+N_{f}$, where $N$, $N_{l}$, and $N_{f}$
are, respectively, the number of vertices, links and faces, is fixed
by the topology so that all planar graphs with the topology of a sphere
share the same value $\chi=2$.
One more constraint, arising from the fact that each link is bounded 
by two vertices and that three links intersect at each vertex, 
reads $2N_{l}=3N$.
These constraints imply that $N_{f}$ and $N_{l}$ are fixed if $N$
is fixed.

\begin{figure}
\includegraphics[clip,width=0.5\textwidth,keepaspectratio]{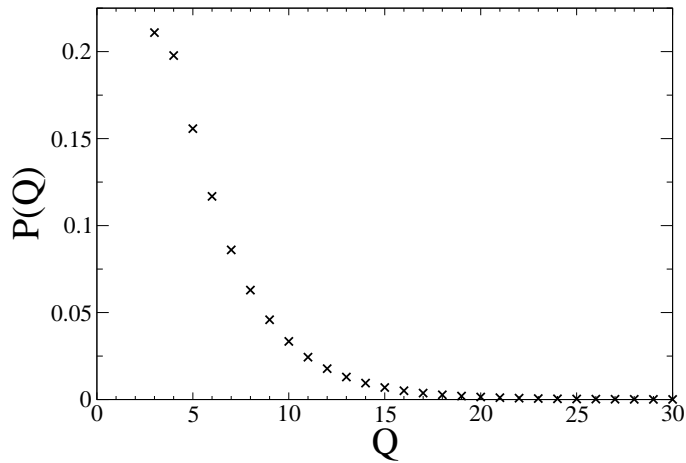}

\caption{Distribution of the size of the faces }

\label{pn_th} 
\end{figure}

By giving each graph $G$ a Boltzmann weight $w(G)$, this class is
turned into a statistical model of random graphs. In this paper, we
choose $w(G)=1$ for each graph. The partition function is given by
\begin{eqnarray*}
Z_{N} & = & \sum_{G\in\Phi^{3}|_{N}}\,\frac{1}{C(G)}\end{eqnarray*}
 where the sum is over all $\Phi^{3}$ graphs with $N$ vertices as
defined above. $C(G)$ is a symmetry factor which avoids the overcounting
of some symmetric graphs and is almost always equal to one for large
graphs. There is an apparent regularity in these graphs. Namely connectivity,
number of vertices, links and faces are fixed so that these graphs
locally look like the honeycomb lattice. Moreover, the average size
of a face, \textit{i.e} the number of links surrounding a face, is
equal to six for large $N$. However, the size of a face is, in fact,
a random variable following a distribution $P(Q)$ exactly known \cite{key-7,key-7BIS}
(see figure \ref{pn_th}), and the correlations associated to the variable $Q$ are
long range, decreasing as $1/r^2$ where $r$ is the geodesic distance \cite{key-6BIS}.
Note that for a face $(i)$ with size $Q_i$, the deviation of 
$Q_i$ from the average value $<Q>=6$ defines the curvature $R_i \propto (Q_i -6)/Q_i$.

This implies that these graphs are very different from the honeycomb
lattice at large distance. For instance, there are objects called baby
universes that induce a self-similar structure for these graphs.
A baby universe is a connected part of a graph linked to it by a boundary 
called a neck. If we call $B$ the size of a baby-universe and $l$ the linear
dimension of its neck, we must have $l \ll B^{1/2}$, {\it i.e.} each baby universe is
connected to the rest of the graph by a small boundary. On trivalent 
graphs, the minimum size of a neck is $l=3$, and baby universes with such necks 
are called minimum neck baby universes (minbus). There are more of minbus than of
the other baby universes and their distribution has been exactly calculated \cite{key-8}.
 Moreover,
baby universes can grow on other baby universes, so that each graph is like a tree made of 
a central part (the root) on which several branches of baby universes are growing.
The global spherical topology ensures that branches do not intersect. This tree
has a self-similar structure that can be seen as follows: consider the family of
baby universes with a given ratio $f=l/\sqrt{B}$ between the linear size $l$ of their neck
and the square root of their size $B$. In a way, $f$ defines a family of baby universes
with the same shape. Then, let $a(B \rightarrow 2 B,f)$ 
be the (average) fraction of the total area of a graph included in baby universes
with a shape characterized by $f$ and with size ranging from $B$ to $2 B$. It 
can be shown that $a(B \rightarrow 2 B,f)$ is independent of $B$. So, if we probe a graph, it looks
similar at any scale, at least in the limit of large graphs.
  
The fractal structure of these graphs described above is characterized by a Hausdorff 
(fractal) dimension $d_{H}=4$ \cite{key-9}.
This dimension is defined by the scaling law \begin{eqnarray*}
\left\langle N_{r}\right\rangle  & \sim & r^{d_{H}}\end{eqnarray*}
 where $\left\langle N_{r}\right\rangle $ is the average number of
vertices at (geodesic) distance lower than $r$ from any arbitrary
vertex.

\section{Targeted attack and robustness of planar $\Phi^{3}$ random graphs\label{sec:Targeted}}

\subsection{Definitions}

A targeted attack is a process that systematically removes highly
connected nodes of a graph. In fact, as each vertex of a $\Phi^{3}$
graph is always trivalent, we define targeted attack as nonrandom
removal of faces - instead of vertices - with large number of edges.
Note that, as $\Phi^{3}$ planar graphs are dual to planar triangulations,
this is equivalent to nonrandom removal of highly connected vertices
of a triangulation.

We introduce a cutoff number $k_{max}$ and define the removal process
as follows: for a given graph $G$, we list all faces whose size -
defined as the number of their edges - is greater than $k_{max}$.
Then, each face in this list is removed, \textit{i.e.} all vertices
and links which belong to its edges are removed. The resulting amputed
graph $\tilde{G}$ is generally not connected. Instead, $\tilde{G}$
consists in a collection of connected smaller planar graphs called
connected clusters. As we consider the statistical ensemble $\Phi^{3}|_{N}$
of planar $\Phi^{3}$ random graphs with $N$ vertices, we are interested
in average properties under targeted attacks, namely the properties
of a graph $\tilde{G}$ associated to a graph $G$ picked at random
in $\Phi^{3}|_{N}$.

Robustness of a graph to targeted attacks means its ability to preserve
its large scale connectivity after a nonrandom deletion of its highly
connected nodes. This can be measured by its diameter, defined as
the average geodesic distance between two arbitrary nodes, \textit{i.e.}
the average length of the shortest path between two arbitrary nodes
\cite{key-2}. Alternatively, as the effect of targeted attacks is
to fragment graphs into smaller connected clusters, robustness can
be measured by the average density of the largest connected cluster
\cite{key-1}, denoted $P$ in this paper. Graphs in $\Phi^{3}|_{N}$
are connected, so that $P$ is equal to one before any targeted attack.
To get more insight on the consequences of targeted attacks, we also
measure the density of the second largest cluster, denoted $s_{2}$,
and the mean size of finite clusters, denoted $S$. The exact definition
of $S$ is the following \cite{key-10} \begin{eqnarray*}
S & = & {\sum_{s}}^{'}s^{2}\ n(s)/{\sum_{s}}^{'}s\ n(s)\end{eqnarray*}
 where $n(s)$ is the average number of clusters made of $s$ vertices
and ${\sum_{s}}^{'}$ means that the largest cluster is excluded from
the sum.

\subsection{Numerical experiment}

\noindent We generated graphs in $\Phi^{3}|_{N}$ following the method
described in appendix \ref{sec:appA}. For each graph in a Monte-Carlo
series, we performed nonrandom removal of faces - for a given value
of $k_{max}$ - and measured the size of the largest connected cluster
by a method described in appendix \ref{sec:appB}. We simulated graphs
in $\Phi^{3}|_{N}$ with $N$ ranging from 1000 to 32000, and the
parameter $k_{max}$ ranging from 8 to 30 for each value of $N$,
except for $N=32000$ where $k_{max}$ was ranging from 12 to 18.
For each couple $(N,k_{max})$, we made 512 measurements separated
by 1000 $N_{l}$ local updating moves (flips) called $T_{1}$ moves 
(see appendix \ref{sec:appB}). Errors were estimated using the
standard jackknife method and error bars are systematically plotted
in the figures.

\subsection{Results}

\begin{figure}
\includegraphics[clip,width=0.5\textwidth,keepaspectratio]{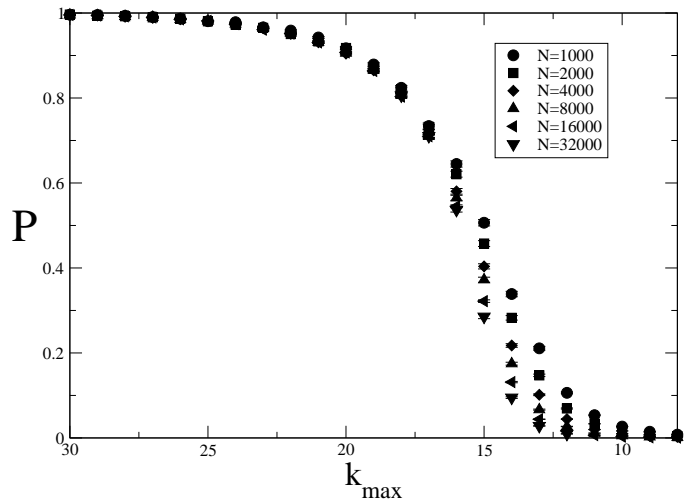}

\caption{Density of the largest cluster versus $k_{max}$.\label{fig:P_vs_k}}
\end{figure}

The density of the largest cluster $P$ is plotted as a function of
the cut-off $k_{max}$ in figure \ref{fig:P_vs_k}. As expected, for
very large $k_{max}$, $P\simeq1$ as graphs in $\Phi^{3}|_{N}$ are
connected. We can see that for values of $k_{max}\gtrsim17$, $P$
slowly decreases with $k_{max}$, independently of the sizes of the
graphs. Then, there is a fast fall of $P$ down to zero. The falling
rate is more and more pronounced when the size is increased. This
means that a targeted attack with $k_{max}$ greater than about 17
does not significantly affect the connectivity of the graphs - \textit{i.e.},
for these values of $k_{max}$, more than $70\%$ of the vertices
belong to the giant cluster. Conversely, $k_{max}$ has to be smaller
than about 13 to completely fragment the graphs into small clusters.

\begin{figure}[h]
\includegraphics[clip,width=0.5\textwidth,keepaspectratio]{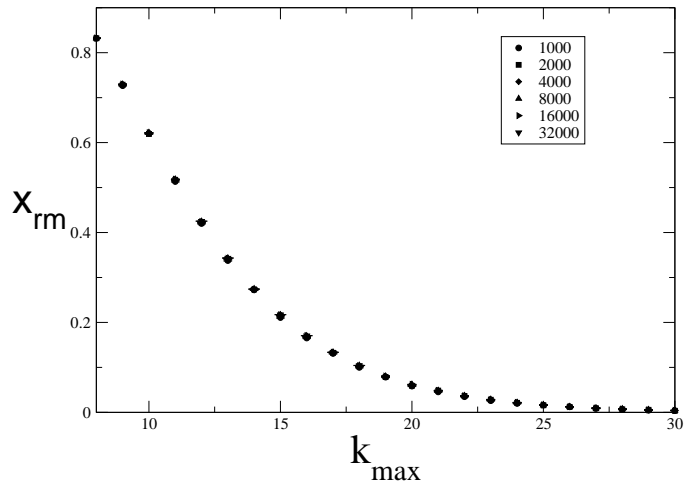}

\caption{Fraction of removed vertices versus $k_{max}$}

\label{x_rm} 
\end{figure}

\begin{figure}[h]
\includegraphics[clip,width=0.5\textwidth,keepaspectratio]{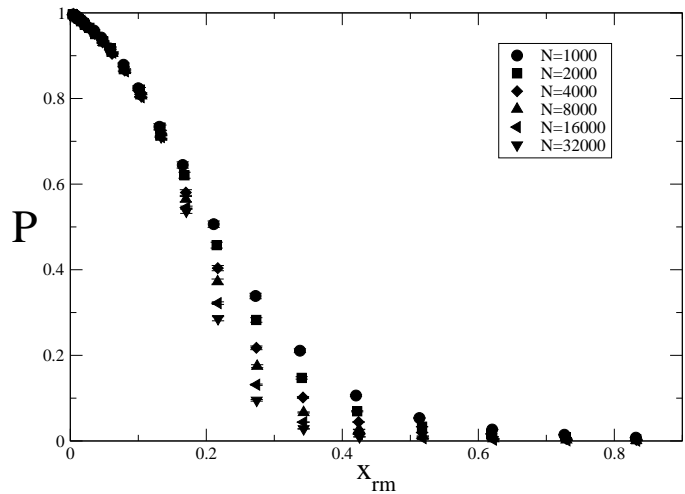}

\caption{Density of the largest cluster versus the fraction of removed
vertices}

\label{fig:P_vs_frac} 
\end{figure}

The average fraction of removed vertices is called $x_{rm}$. It is
a smooth decreasing function of $k_{max}$ as can be seen in figure
\ref{x_rm}. It is interesting to plot $P$ against $x_{rm}$. This
is shown in figure \ref{fig:P_vs_frac}.

We can see that a targeted attack significantly affects the connectivity
only when a large fraction of the vertices are removed. For example,
$P\gtrsim0.7$ for $x_{rm}\lesssim0.13$ and $P\gtrsim0.5$ for $x_{rm}\lesssim0.17$.
In this sense, we can say that planar $\Phi^{3}$ random graphs are
rather robust to targeted attacks.

\begin{figure}[h]
\includegraphics[width=0.5\textwidth,keepaspectratio]{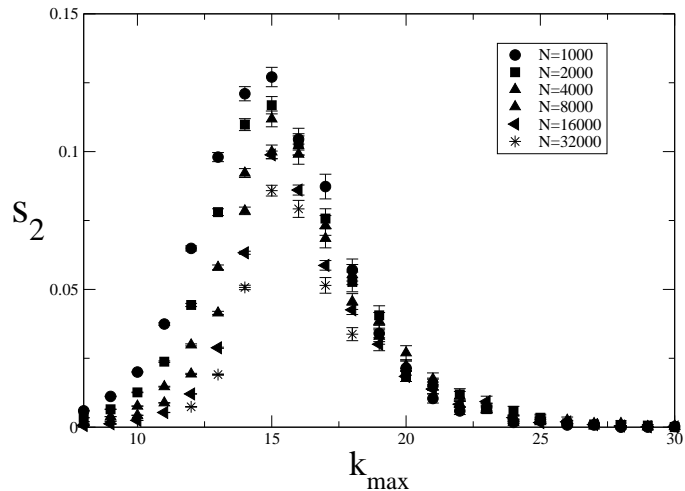}

\caption{Density of the second largest cluster versus $k_{max}$}

\label{s2} 
\end{figure}

\begin{figure}
\includegraphics[clip,width=0.5\textwidth,keepaspectratio]{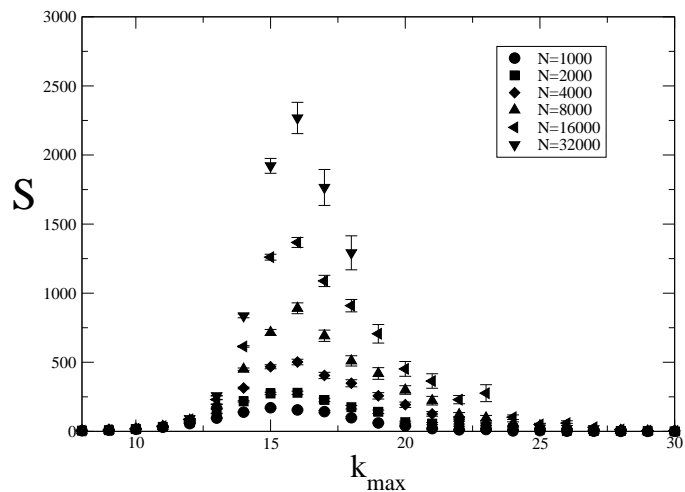}

\caption{Average size of finite clusters versus $k_{max}$}

\label{smean} 
\end{figure}

The analysis of the density of the second largest cluster $s_{2}$
and of the average size of finite clusters $S$, plotted in figures
\ref{s2} and \ref{smean} respectively, gives more information on
the mechanism of fragmentation of graphs under targeted attacks. For
large $k_{max}$, $s_{2}\simeq 0$, meaning that most of vertices belong
to the largest cluster. Then, $s_{2}$ sharply increases when $k_{max}$
decreases and eventually reaches a peak for $k_{max}\simeq15$. This
is due to a small gradual decrease of the connectivity, \emph{i.e.}
a small fraction of vertices get disconnected from the largest cluster.
However, this phenomenon is marginal as $s_{2}$ does not exceed 0.15
and $S$ is not greater than about 8\% of the size. So, in this region
$k_{max}\gtrsim15$, the largest cluster completely dominates with
only a few number of much smaller clusters. This is corroborated by
the behavior of $S$ in this region. As the largest cluster is excluded
from this quantity, $S\simeq 0$ for very large $k_{max}$. As $k_{max}$
decreases, $S$ is essentially influenced by the second, third, etc,
largest clusters and so it increases and reaches also a peak. Then,
when $k_{max}$ continues to increase, \textit{i.e.} when $s_{2}$
and $S$ go through their peak, there is a dramatic change in the
fragmentation process. $P$ undergoes a fast drop, meaning that there
is an acceleration in the fragmentation and more and more vertices
get disconnected from the initial largest cluster. The behaviors of
$s_{2}$ and $S$ are similar. This is due to the fact that more and
more small clusters appear, so that their average size is small, even
for the largest, second largest,third largest, etc, clusters.

\section{Non-uniform percolation\label{sec:Non-uniform}}

\subsection{Percolation threshold}

The fragmentation process described above is very similar to the  percolation
mechanism \cite{key-10}. This is, in fact, expected as nonrandom removal
of nodes under targeted attack is equivalent to percolation with non-uniform
occupation probability \cite{key-1}. In the case studied
here, a vertex is occupied if and only if it does not belong to an
edge of a face whose size is greater than $k_{max}$. The behavior
of $P$, $s_{2}$ and $S$ can then be interpreted by saying that
there exists a critical value $k_{c}\simeq 15$ so that percolation
takes place for $k\ge k_{c}$. In the following, we consider $k_{max}$
as a continuous parameter; more precisely, it is as if the observables
$P$, $s_{2}$ and $S$ were defined for all real (positive) values
of $k_{max}$ but that numerical data were available only for integer
values of $k_{max}$. Note that $k_{max}$ can be seen as a cutoff
on the local scalar curvature. By taking an appropriate continuum
limit when $N\rightarrow\infty$ , the curvature is, of course, a continuous
observable so that a cutoff on curvature has to be continuous in
this limit. In other words, in any process of coarse graining, $k_{max}$
would effectively become a continuous parameter.

In order to get a more precise value of $k_{c}$, we define a finite
size critical parameter $k_{c}(N)$ as the value of $k_{max}$ for
which $s_{2}$ is maximum for $N$ fixed. By fitting the data of $s_{2}$
with a Gaussian curve, we obtained $k_{c}(N)$. The result is shown
in figure \ref{kcN}. The behavior of $k_{c}(N)$ can be estimated
by standard finite size scaling analysis (see appendix \ref{sec:appC}).
This allows us to predict the behavior of $k_{c}(N)$ approaching
$k_{c}$ as $k_{c}(N)-k_{c}\sim N^{-1/\nu d_{H}}$ with $\nu>0$ a
constant explained in section \ref{sub:Critical-exponents}. So, we
fitted the values of $k_{c}(N)$ with the law $k_{c}(N)=k_{c}+c\ N^{-1/\nu d_{H}}$,
with $c$ a constant. We obtained $k_{c}=15.8(2)$.

\begin{figure}[h]
\includegraphics[clip,width=0.5\textwidth,keepaspectratio]{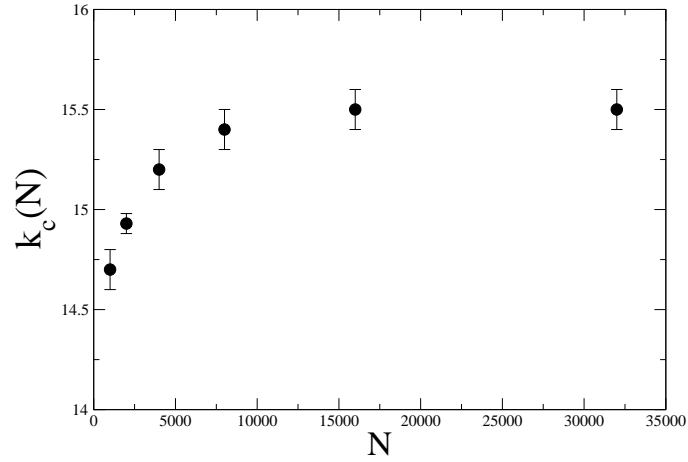}

\caption{Finite size critical parameter $k_{c}(N)$}

\label{kcN} 
\end{figure}

\subsection{Critical exponents $\nu$, $\beta$ and $\gamma$ \label{sub:Critical-exponents}}

The point of view of non-uniform percolation with a percolation threshold
$k_{c}$ strongly suggests that the observables $S$ and $s_{2}$
should obey the following scaling laws for $k_{max}$ near $k_{c}$
\cite{key-10,key-11} (recall that we consider here $k_{max}$ as
a continuous parameter) \begin{eqnarray*}
s_{2} & \sim & \left(k_{max}-k_{c}\right)^{\beta}\\
S & \sim & \left(k_{max}-k_{c}\right)^{-\gamma}\end{eqnarray*}

Their finite size counterparts are 
\begin{eqnarray*}
s_{2}(N) & \sim & N^{-\beta/\nu d_{H}}\: F\left[\left(k_{max}-k_{c}\right)\, N^{-1/\nu d_{H}}\right]\\
S(N) & \sim & N^{\gamma/\nu d_{H}}\: G\left[\left(k_{max}-k_{c}\right)\, N^{-1/\nu d_{H}}\right]\end{eqnarray*}

with $F$ and $G$ two scaling functions. The exponent $\nu$ characterizes
the correlation length associated with a percolation transition. In
particular, this exponent controls the way $k_{c}(N)$ tends to its
infinite size value $k_{c}$ . So, the fit performed in the previous section
also allows us to extract the value of $\nu$. The result is $1/\nu d_{H}=0.4(2)$.
In order to extract the values of exponents $\beta$ and $\gamma$,
we have measured the maximum of $s_{2}$ and $S$ , which should scale
as
\begin{eqnarray*}
s_{2}^{max}(N) & \sim & N^{-\beta/\nu d_{H}}\\
S^{max}(N) & \sim & N^{\gamma/\nu d_{H}}\end{eqnarray*}

\begin{figure}[h]
\includegraphics[clip,width=0.5\textwidth,keepaspectratio]{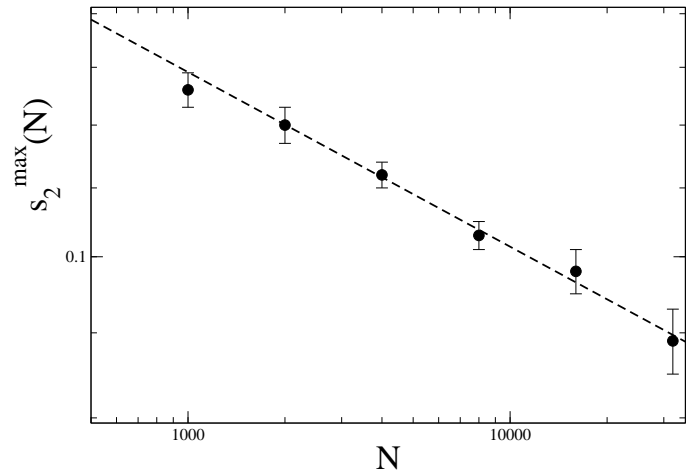}

\caption{Maximum of density of the second largest cluster versus $N$,
in logscale. The dotted line is the best fit\label{fig:s2max}}
\end{figure}

\begin{figure}[h]
\includegraphics[clip,width=0.5\textwidth,keepaspectratio]{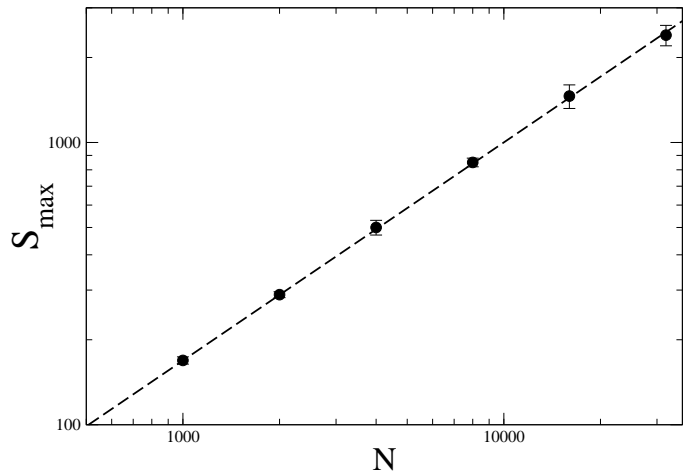}

\caption{Maximum of the average cluster size versus $N$. The dotted
line is the best fit\label{fig:Smax}}
\end{figure}

The results are shown in figures \ref{fig:s2max} and \ref{fig:Smax}.
We fitted the data and obtained $\beta/\nu d_{H}=0.105(5)$ and $\gamma/\nu d_{H}=0.77(1)$.

\subsection{Systematic errors}

Because of the discrete nature of the parameter $k_{max}$, there
are no available data for non-integer values of this parameter. This
inevitably induces systematic errors - in particular, values of $k_{max}$
for which observables are maximum - certainly underestimated in the
standard analysis used here.

\subsection{Comparing with uniform percolation exponents}

Critical exponent are known exactly for uniform percolation on planar
$\Phi^{3}$ random graphs \cite{key-5}, $\beta/\nu d_{H}=0.125$,
$\gamma/\nu d_{H}=0.75$ and $1/\nu d_{H}=0.25$. It is tempting to
compare them with the values found in this work for non-uniform percolation
$\beta/\nu d_{H}=0.105(5)$, $\gamma/\nu d_{H}=0.77(1)$ and $1/\nu d_{H}=0.4(2)$.
In view of systematic errors underestimated, it appears that values
of exponents $\beta$ and $\gamma$ are roughly compatible for uniform
and non-uniform percolation. The exponent $\nu$ seems very different
from the exact value of uniform percolation. However, if we compare
it with the value $1/\nu d_{H}=0.489(9)$ found in numerical simulations
\cite{key-6}, it is once again roughly compatible. In fact, as in
uniform percolation, the fractal dimension $d_{H}$ is very sensitive
to finite size effects. However, measures done in this work are not
precise enough to conclude that critical exponents are the same for
uniform and non-uniform percolation for this problem. It is not excluded
either and this would imply a kind of universality between uniform
and non-uniform percolation on these graphs.

\section{Conclusions\label{sec:Conclusions}}

In the first part of this work, the behavior of planar $\Phi^{3}$
random graphs under targeted attacks is investigated. The method consists
in removing all vertices that belong to an edge of a face whose size
is greater than a parameter $k_{max}$. It appears that this family
of graphs is rather robust to this nonrandom removal as $k_{max}$
has to be rather small ($k_{max}\simeq15$ ) to significantly affect
the global connectivity of these graphs, measured by the density of
the largest cluster. This threshold corresponds to the removing of
about $20$\% of the vertices. Note that the case of random failure
is equivalent to uniform site percolation. This problem has already
been investigated and the high value of the percolation threshold
means that these graphs are rather fragile against random failure
of nodes.

In a second part, we consider the targeted attacks problem as non-uniform
percolation. By measuring two observables characterizing percolation
clusters, namely the density of the second largest cluster $s_{2}$
and the mean size of connected clusters $S$, and by extrapolating
values of $k_{max}$ to non-integer values, we extract a critical
value $k_{c}=15.8(2)$ that can be interpreted as a percolation threshold
for targeted attacks. It is also possible to look for scaling laws
of observables in the vicinity of $k_{max}$ , by analogy to the uniform
percolation problem. This allows us to measure (effective) standard critical
exponent $\beta$, $\gamma$ and $\nu$. Taking into account underestimated
systematic errors due to the fact that $k_{max}$ is, in fact, a discrete
parameter, critical exponents found here appear roughly compatible
with their counterparts of uniform percolation. This is a very interesting
fact as it would show that non-uniform and uniform percolations are
in the same class of universality. If true, this property could be
a result of the hierarchical nature of these graphs that makes them
look like trees of baby universes (BUs).

Indeed, consider a vertex in a BU. For uniform percolation, a necessary condition 
for this vertex to belong to a giant cluster is that at least one vertex in the neck
of the BU belongs to this giant cluster. The probability for this condition
is rather small as necks are very small regions of the graph, so that the probability
for a vertex to belong to a giant cluster is smaller if the vertex is in a BU.
 Consider now the case of 
targeted attacks. The probability for a vertex to be removed is, in fact, greater if it 
is in a BU as, on average, there is more curvature in BUs. Then, the probability that
a vertex is in the boundary of a large face is greater if the vertex is in a BU. So, 
in both cases (uniform percolation and targeted attacks), the structure of BUs induces
the same effect: the probability to be occupied is smaller for a vertex in a BU.    
However, this conjecture of universality between uniform and non-uniform
percolation has to be more extensively and precisely investigated. 

\appendix

\section{Generating and sampling $\Phi^{3}$ graphs\label{sec:appA}}

\begin{figure}[h]
\includegraphics[clip,width=0.5\textwidth,keepaspectratio]{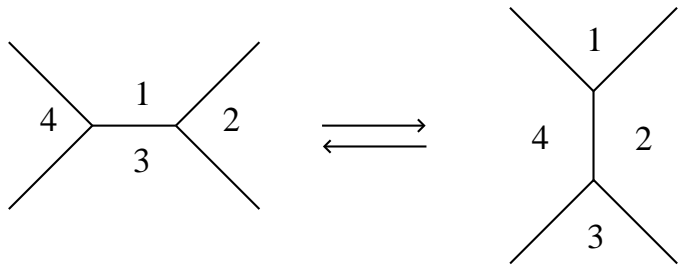}

\caption{$T_{1}$ move involving faces 1,2,3 and 4.\label{fig:T1}}
\end{figure}

The Monte-Carlo sequence starts with a graph in $\Phi^{3}|_{N}$ randomly
generated as follows: we start from a tetrahedron and, then, add vertices
one by one in randomly chosen faces (triangles). Each new vertex is
linked to the vertices of the corresponding triangle. The process
is repeated until we obtain a graph (polyhedron) with $N$ triangles.
This graph becomes a $\Phi^{3}$ graph - denoted $G_o^{N}$ - by duality,
 \textit{i.e.} by replacing
each triangle of the polyhedron by a vertex linked to the vertices
replacing the adjacent faces of the initial triangle. The resulting
graph is topologically equivalent to a sphere with $N$ trivalent
vertices. Then, a series of graphs is obtained by using standard flips
of links ($T_{1}$ moves shown in figure \ref{fig:T1}) performed
on randomly chosen links. It has been shown \cite{key-7} that $T_1$
moves define an ergodic transformation in the ensemble $\Phi^{3}|_{N}$,
so that whatever the starting graph $G_o^{N}$, we obtain a nonbiased
sampling with this method. 

\section{Cluster structure\label{sec:appB}}

In order to measure the connected cluster structure of graphs after
a targeted attack, we use a breadth-first search algorithm similar
to the Wolff algorithm \cite{key-12}. It recursively constructs all clusters
for a given graph as follows: at {}``time'' $n$, clusters $c_{1},c_{2},\ldots,c_{n}$
have already been detected. All sites in these clusters are labeled
\char`\"{}visited\char`\"{}. At time $n+1$, a not yet visited site
(called $v_{o}$) is chosen. By definition, it does not belong to
any cluster already detected. $v_{o}$ is the root of the new cluster
$c_{n+1}$; it is then labeled \char`\"{}visited\char`\"{} and put
in a (empty) list $Q$. The following procedure is now recursively
applied to $Q$ : for each site $v$ in $Q$, all not yet visited
neighbors of $v$ (on the graph) are added to $c_{n+1}$, labeled
\char`\"{}visited\char`\"{} and put in $Q$ whereas $v$ is removed
from $Q$ . The procedure stops when $Q$ is empty. Then, cluster
$c_{n+1}$ is completely constructed. The algorithm stops when all
sites have been visited.

\section{Finite size scaling\label{sec:appC}}

Finite size scaling analysis supposes that finite size corrections
to scaling laws near a critical parameter $k_{c}$ are encoded by
scaling functions depending on the ratio between the linear size $L$
of the system and the correlation length $\xi$ . When $L\gg\xi$
, finite size effects should not affect the system and scaling laws
undergo no correction. On the contrary, when $\xi$ and $L$ become
comparable, finite size corrections of scaling laws are expected.
For an observable $O$ depending on a parameter $k$, an infinite
size scaling law $O\sim\left(k-k_{c}\right)^{-z}$ can be rewritten
$O\sim\xi^{z/\nu}$ as the correlation length scales as $\xi\sim\left(k-k_{c}\right)^{-\nu}$.
In particular, $\xi\sim L$ implies $O\sim L^{z/\nu}$. This can be
summarized by the following law: $O\sim L^{z/\nu}\, F\left(\left(k-k_{c}\right)\, L^{1/\nu}\right)$
with $F(x)$ a (scaling) function of the dimensionless ratio 
$x=\left(k-k_{c}\right)L^{1/\nu}\sim\left(L/\xi\right)^{1/\nu}$.
In order to interpolate between the cases $L\gg\xi$ and $L\sim\xi$,
$F(x)$ must verify $F(x)\rightarrow 1$ for $x\sim 1$ and $F(x)\rightarrow x^{-z}$
for $x\gg 1$. Finite size scaling is a powerfull tool for extracting
critical exponents by studying the variations of observables with
the size of the system. This also provides a natural finite size critical
parameter $k_{c}(L)$ : suppose that $F(x)$ reaches a maximum for
$x=x_{o}$. Then, for fixed $L$, the value of $k$ giving a maximum
for $O$ obeys the following relation $\left(k-k_{c}\right)\, L^{1/\nu}=x_{o}$
or, equivalently, $k_{c}(L)$ approaches $k_{c}$ when the size becomes
infinite as $k_{c}(L)-k_{c}\sim L^{-1/\nu}$ \cite{key-10}. However,
graphs in $\Phi^{3}|_{N}$ have no explicit linear size, but the quantity
$N^{1/d_{H}}$, where $d_{H}$ is the Hausdorff dimension, plays this
role so that finite size scaling laws are $O\sim N^{z/\nu d_{H}}\, F\left(\left(k-k_{c}\right)\, N^{1/\nu d_{H}}\right)$
and $k_{c}(N)$ approaches $k_{c}$ as $k_{c}(N)-k_{c}\sim N^{-1/\nu d_{H}}$.

\end{document}